\documentclass{article}

\usepackage[utf8]{inputenc} 
\usepackage[T1]{fontenc}    
\usepackage{hyperref}       
\usepackage{url}            
\usepackage{booktabs}       
\usepackage{amsfonts}       
\usepackage{amsmath}
\usepackage{amssymb}
\usepackage{float}
\usepackage{nicefrac}       
\usepackage{microtype}      
\usepackage{graphicx}
\usepackage{svg}
\usepackage{placeins}
\usepackage{xcolor} 
\usepackage{doi}

\usepackage[backend=biber, sorting=none, style=phys]{biblatex}

\usepackage{mathtools}
\usepackage{caption,subcaption}
\usepackage{fancyhdr,fancybox}
\usepackage[verbose=true,letterpaper,top=1in,bottom=1in,left=0.4in,right=0.3in]{geometry}
\usepackage{authblk}
\usepackage{arxiv}
\usepackage[export]{adjustbox}

\title{Spontaneous spatial sorting by cell shape in growing colonies of rod-like bacteria
}

\author[1]{Mateusz Ratman\textsuperscript{*}}
\author[1]{Jimmy Gonzalez Nu\~nez\textsuperscript{*}}
\author[1]{Daniel A.~Beller}

\affil[1]{Department of Physics and Astronomy, Johns Hopkins University, Baltimore, MD 21211}

\begin{document}
\maketitle

\def\thefootnote{*}\footnotetext{These authors contributed equally to this work.}

\begin{abstract}
     Mechanical interactions among cells in a growing microbial colony can significantly influence the colony's spatial genetic structure and, thus, evolutionary outcomes such as the fates of rare mutations. Here, we computationally investigate how this spatial genetic structure changes as a result of heritable phenotypic variations in cell shape.
     By modeling rod-like bacterial cells as lengthening and dividing circo-rectangles in a 2D Brownian dynamics framework, we simulate the growth of a  colony containing two populations with different aspect ratios.     Compared to monodisperse colonies, such bidisperse colonies exhibit diminished  intermixing between sub-populations when the less elongated cells are too short to nematically order, instead forming large clusters. 
     We find that the cells with longer aspect ratio gradually segregate to the colony periphery. We present evidence  that this demixing is related to nematic order in the bulk and to active nematic mixing dynamics near the periphery. These findings are qualitatively robust across different growth rate protocols and initial conditions. Because the periphery is often an advantageous position when nutrients are limited, our results suggest a possible evolutionary selective pressure of mechanical origin that favors large cell aspect ratio. 
\end{abstract}


\twocolumn
\newpage

\section*{Introduction\label{sec:intro}}

Biological populations commonly exhibit significant spatial structure in their distributions of genotypes, with often profound consequences for their evolutionary dynamics \cite{Hallatschek2007, Excoffier2008, Excoffier_2009, Korolev2010, Farrell2013, Gralka2016, Farrell2017, Gralka2019, Hallatschek2010,Chu_2019,nunez2024range, Wilkins2002,WilkinsonHerbots1998,  Beller2018,}. 
Growing colonies of asexually reproducing, immotile microbes provide model systems for quantitative laboratory studies of spatial structure \cite{Hallatschek2007,Farrell2013,Farrell2017, giometto2018physical,vandenBerg2024}. Colony growth promotes genetic demixing by which an originally well-mixed colony spontaneously segregates its genotypes, in regions of new growth, into clonal sectors with low local genetic diversity \cite{Hallatschek2007, Korolev2010}. 

Mechanical interactions between neighboring cells in the densely packed colony play important and sometimes subtle roles in the emergent spatial genetic structure. For example, a study of growing \textit{Saccharomyces cerevisiae} yeast colonies found that mechanical interactions between competing strains of different fitness favored the slower-growing strain, effectively weakening natural selection \cite{giometto2018physical}.  In simulations of rod-shaped bacteria, mechanical interactions were sufficient to explain an experimentally observed transition between circular and branching colony growth morphologies \cite{Farrell2013}.  Mechanical interactions with secreted extracellular polymeric substances may also help bacteria to maintain compact clusters of cells \cite{Ghosh2015}.

However, little is known about  how spatial genetic structure is modified by mechanical interactions between cells with differing shape phenotypes, and how this contribution to population structure may thus produce selective pressures favoring certain cell shapes.  Bacteria regulate their cell shape with high precision under normal growth conditions
 \cite{koch2001bacterial, wang2010robust, amir2014cell, cesar2017thinking}, yet remarkable diversity in cell shape exists among  bacterial species \cite{Yang2016} as well as yeast species \cite{Chavez2023}, influenced by a variety of selective pressures.


Nematic orientational order, exhibited more strongly for cells of higher aspect ratio, is a key mechanism by which cell shape influences larger-scale structure. For example, in Ref.~\cite{vandenBerg2024}, alignment of longer cells in nematic bands is important to their domination of the colony periphery. Interestingly, recent experiments showed that large cell aspect ratio confers a selective advantage to bacterial colonies growing in three-dimensional porous media \cite{Sreepadmanabh2024}.

While swarms of motile bacteria have provided a paradigmatic example of active nematics  \cite{copenhagen2021topological,Yashunsky2024}, growing colonies of immotile microbes with sufficiently high cell aspect ratio represent a distinct class of active nematic \cite{Volfson2008, Farrell2017,arxiv_vanHolthe,Farrell2013,DellArciprete2018, Doostmohammadi2016,You2021,You2018,Sengupta2020}, exhibiting characteristics peculiar to \textit{proliferating} active matter systems of ever-increasing mass \cite{Hallatschek2023}. Even though proliferating active nematics lack the defect-driven chaotic mixing of typical active nematics \cite{decamp2015orientational, tan2019topological}, nematic order is still correlated with self-mixing flows that hinder growth-induced genetic demixing~\cite{Schwarzendahl2022}. 

The evolutionary consequences of proliferating active nematic dynamics in colonies of coexisting cell shapes have been so far unexplored. To the extent that maintaining local diversity in genotypes is advantageous, Ref.~\cite{Schwarzendahl2022} identified a possible selective pressure favoring high cell aspect ratios through active nematic mixing. Such a pressure would necessarily act through variation in cell shape phenotype, which was not investigated in that prior work. In general, phenotypic diversity is common even among genotypically identical populations \cite{Spudich1976} and can represent a survival strategy through bet-hedging~\cite{Veening2008}.

In this work, we computationally investigate how diversity in cell shape phenotypes influences genetic spatial structure within a growing microbial colony through mechanical interactions alone. We employ an agent-based, overdamped molecular dynamics approach to model bidisperse bacterial colonies, whose cells have heritable aspect ratios that differ between two coexisting types. Through spatial measures of population genetic diversity and active nematic ordering, we identify a regime in which active mixing is suppressed and genetic segregation is enhanced. Strikingly, the periphery of the expanding colony tends to be dominated by the cell type of larger aspect ratio. This spontaneous, partial shape-sorting occurs through two separate mechanisms: Longer cells are more likely to be driven toward the periphery by growth-induced forces, and once at the periphery, they have a higher propensity to remain there due to  tangential alignment. Alongside active nematic dynamics, our results indicate the potential importance of an active analogue to granular convection, best known through the Brazil nut effect \cite{koda1996smectic, adams1998entropically}. 
These complex dynamics arising from purely mechanical interactions may produce important selective pressures on cell shape in scenarios of limited nutrients \cite{Farrell2017}, bet-hedging \cite{Veening2008}, and mutualistic or antagonistic interactions between species \cite{Bronstein2003, Korolev2011}.


\section*{Model\label{sec:model}}

\subsection*{Hard Rod Model for Bacterial Growth}

\begin{figure}[htb]
    \centering
    \includegraphics[width=\linewidth]{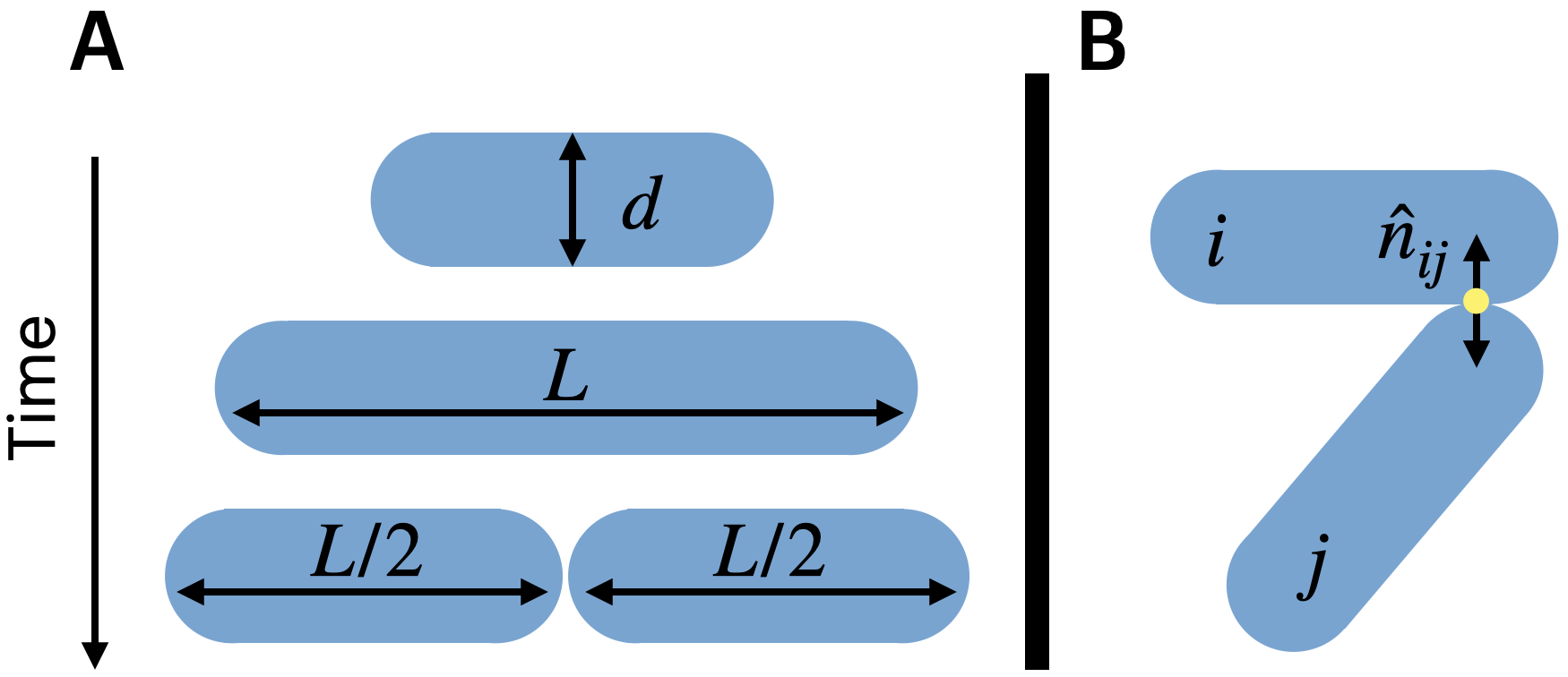}
    \caption{Schematic of model for rod-like bacterial cells as lengthening circo-rectangles. (A) A single cell of width $d$ lengthens until it reaches a length $L$ and splits into two cells of equal length $L/2 $. (B) Overlapping cells repel with forces directed along $\hat n_{ij}$, the normal vector at the approximate contact point. }
    \label{fig:growth_process_model}
\end{figure}

We use a modified version of an agent-based computational model used in  previous mechanical studies of bacterial growth~\cite{Farrell2013, Farrell2017, You2021, Schwarzendahl2022}  to simulate a bacterial colony growing on a surface and containing two cell shape phenotypes. Specifically, each phenotype corresponds to a circo-rectangle shape with one of two maximum lengths $L_X$, $X\in \{A,B\}$, giving two maximum aspect ratios $a_X = L_X / d$ where the width $d$ is the same for both phenotypes and is held constant. We will refer to these two groups of cells as Population A and Population B. Cells increase their length at a constant elongation rate $g_X$ and divide into two cells of length $L_X / 2$ upon reaching their maximum length $L_X$. Additionally, upon division, the two resulting cells are given new elongation rates that are uniformly selected from the range $[ g_X/2, 3 g_X / 2]$. This procedure for elongation rates is applied in order to mitigate artifacts due to synchronization of division times \cite{Schwarzendahl2022,You2018,Langeslay2023}.

The dynamics of each cell follows over-damped motion governed by the following  equations, for the center of mass position $\vec r_i$ and long-axis orientation $\theta_i$ of cell $i$:
\begin{align}
    \frac{d \vec{r}_i}{d t} &= \frac{1}{\zeta l_i} \sum_j \vec{F}_{ij} \\
    \frac{d \theta_i}{d t} &= \frac{12}{\zeta l_i^3} \sum_j \left( \vec{r}_{ij} \times \vec{F}_{ij} \right) \cdot \hat{e}_z.
\end{align}
Here, $\zeta$ is a friction coefficient, $\hat{e}_z$ is the unit vector perpendicular to the cells' plane of motion, $\vec{r}_{ij} = \vec{r}_i - \vec{r}_j$ is the separation vector between the positions of cell $i$ and cell $j$, $l_i$ is the $i$th cell's current length, and $\vec{F}_{ij}$ are the forces between cells. In our model, inter-cell forces are modeled by Hertzian contact theory, 
\begin{equation}
    \vec{F}_{ij} = F_0 d^{1/2} h_{ij}^{3/2} \hat{n}_{ij},
\end{equation}
where $F_0$ is the strength of repulsion, $h_{ij}$ is the overlap between two cells, and $\hat{n}_{ij}$ is the vector normal to each cell surface at the point of closest contact between cell $i$ and cell $j$. Schematic illustrations for the growth process and cell-cell interactions are shown in Fig.~\ref{fig:growth_process_model}.

Our model assumes that nutrients are always uniformly distributed, making growth rates independent of position. This bulk growth scenario is in contrast to limited-nutrient scenarios in which growth is limited to cells near the colony periphery \cite{Volfson2008,Gralka2016, Hallatschek2007,Farrell2013}.
We examine two types of growth rate rules, in which the cell types have either  (1) equal division time $T_A = T_B$, where $T_X = 0.5L_X/g_X = 2 \times 10^6$, resulting in an approximately constant ratio of cell numbers; or (2) equal elongation rate $g_A = g_B$.
Colonies are initialized as isotropic droplets of 25 cells of each type with uniformly distributed orientations and positions, and are grown to a size of 25000 cells. Details of the initialization protocol are given in~\textbf{Materials and Methods}.

To keep our parameter space tractable, we set the friction coefficient $\zeta$ by fixing the dimensionless parameter $g_X\zeta / F_0 L_X = 5 \times 10^{-7}$, the same value used in other recent studies employing a similar model \cite{You2018,Schwarzendahl2022,Langeslay2023}.

\section*{Results\label{sec:results}}
\subsubsection*{Spatial Sorting During Equal Division-Time Growth}

\begin{figure*}
    \centering
    \includegraphics[width=0.9\linewidth]{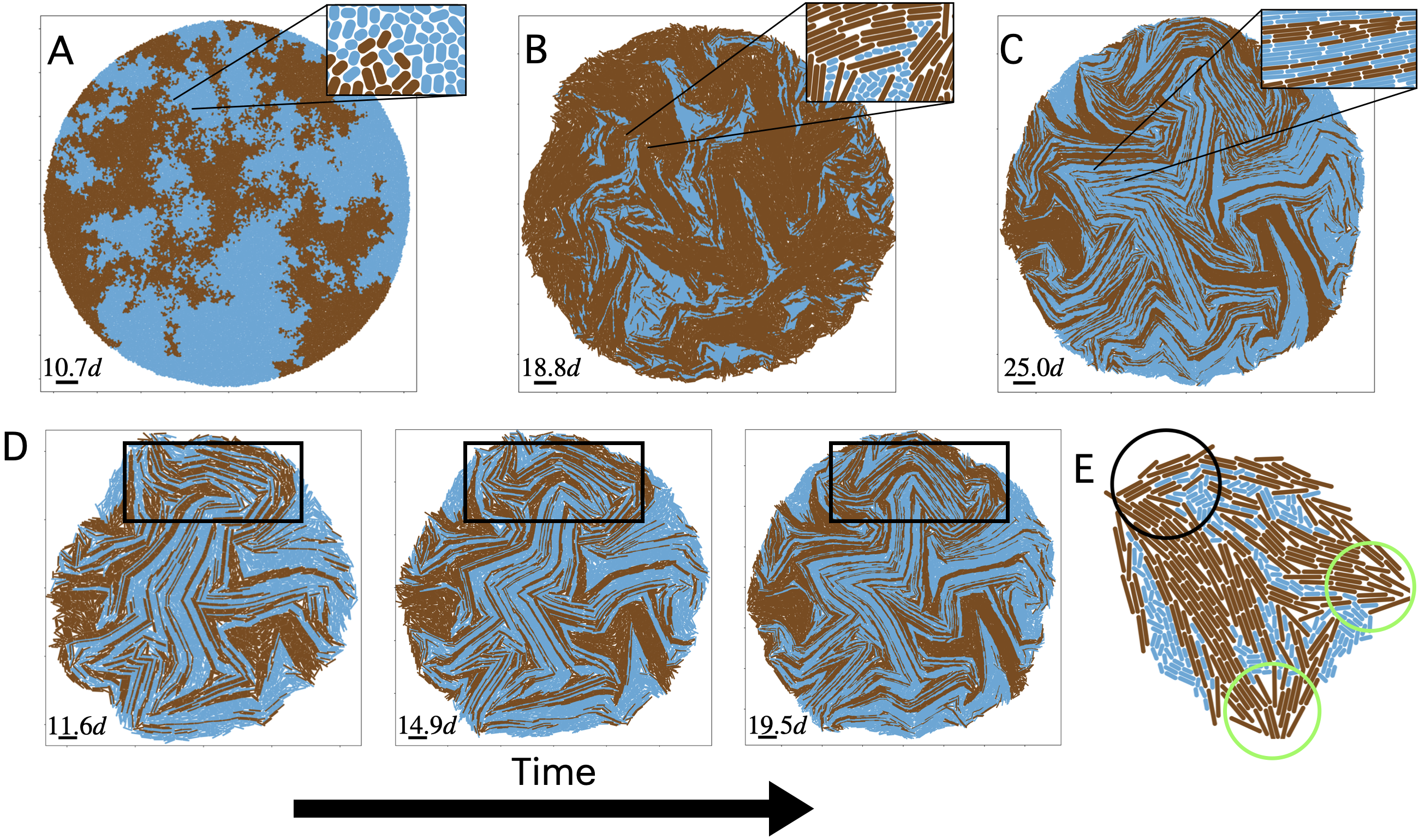}
    \caption{Snapshots of simulated colonies with approximately 25000 cells of two cell types A (blue) and B (brown), which have aspect ratios $(a_A, a_B)$ of \textbf{(A)} $(2,2)$, \textbf{(B)} $(2, 10)$, \textbf{(C)} $(10, 10)$. $\textbf{(D)}$ Sequence of snapshots for colony shown in (C) at times $t=6.6T$, $t=7.4T$ and $t=8.2T$. A region that exemplifies high  active nematic mixing is highlighted by black boxes. \textbf{(E)} Snapshot of a small colony of $ 464 $ cells with aspect ratios $a_A = 4$ (blue) and $a_B=10$ (brown). 
    Population B cells found at the colony periphery  preferentially orient tangentially, such as in the sub-region marked by the black circle. At other locations on the periphery where Population B cells are radially oriented, indicated by green circles, these cells tend to be part of a radially oriented nematic domain protruding out from the interior. The two cell types have equal division times $T_A=T_B$.}
    \label{fig:growth}
\end{figure*}

We first consider the growth rate rule (1) of equal division times $T_A=T_B$. Figure~\ref{fig:growth}A,B,C show final-time snapshots of colonies simulated from an initially isotropic droplet containing equal numbers of the two cell types, for various combinations of aspect ratio. Videos of the growth are shown in Supplementary Movies S1-3 in the \textbf{SI Appendix}. As was seen in Ref.~\cite{Schwarzendahl2022},  small aspect ratio cells ($a_A = a_B = 2$, Fig.~\ref{fig:growth}A) produce larger regions of single cell type compared to high aspect ratio cells ($a_A = a_B = 10$, Fig.~\ref{fig:growth}C). The latter exhibit to some extent the buckling and self-mixing dynamics characteristic of active nematics \cite{Keber_2014, tan2019topological},  evidenced by large-scale bands of alternating color in Fig.~\ref{fig:growth}C.

 In Fig.~\ref{fig:growth}B, we show a bidisperse colony with cells of maximum aspect ratio $a_A =2$ (light blue) and $a_B = 10$ (brown). The cells of smaller aspect ratio become isolated in pockets between nematically ordered regions of high-aspect-ratio cells (Fig.~\ref{fig:growth}B inset). Strikingly, we observe that bidisperse colonies undergo spontaneous  spatial sorting of cell phenotypes such that cells of higher aspect ratio occupy an increasingly large proportion of the colony periphery over time. This sorting is surprising in light of the active mixing also seen, whereby buckling of nematically aligned domains of longer cells causes stretching of the domains of short cells, and thus enhanced intermixing of the two cell types in the bulk. A region of a colony with significant active mixing dynamics is highlighted in Fig.~\ref{fig:growth}D. While nutrients are uniformly abundant in our simulations, under nutrient-scarce conditions the colony periphery may be the only place where growth occurs, implying a potential selective pressure in favor of traits that allow dominance of the periphery.
 
 To understand this interplay of mixing in the bulk and sorting at the periphery, we examine the orientations of cells at the periphery. Fig.~\ref{fig:growth}E shows a snapshot of the colony with $a_A = 4$ and $a_B = 10$ at an early time. Large-aspect-ratio cells at the colony periphery tend to align tangentially, such as the brown rods in the example highlighted by the black-circle annotation in  Fig.~\ref{fig:growth}E. We hypothesize that one factor contributing to the dominance of the periphery by high-aspect-ratio cells is these cells' greater likelihood, when tangentially oriented at the periphery, to encounter a gap between nearby cells large enough to allow them to enter the bulk, in a reference frame co-moving with the traveling periphery. This means that a phenotype with larger aspect ratio  will have a greater tendency to persist on the boundary after occupying some portion of it. On the other hand, a minority of longer cells at the periphery are not  oriented tangentially; two examples of near-normal alignment are highlighted by green circles in Fig.~\ref{fig:growth}E. Such cells tend to be connected to nematically aligned domains protruding out from the bulk, a structure that we address below in \textbf{Radial Order}. 

\begin{figure*}
    \centering
    \includegraphics[width=0.9\linewidth]{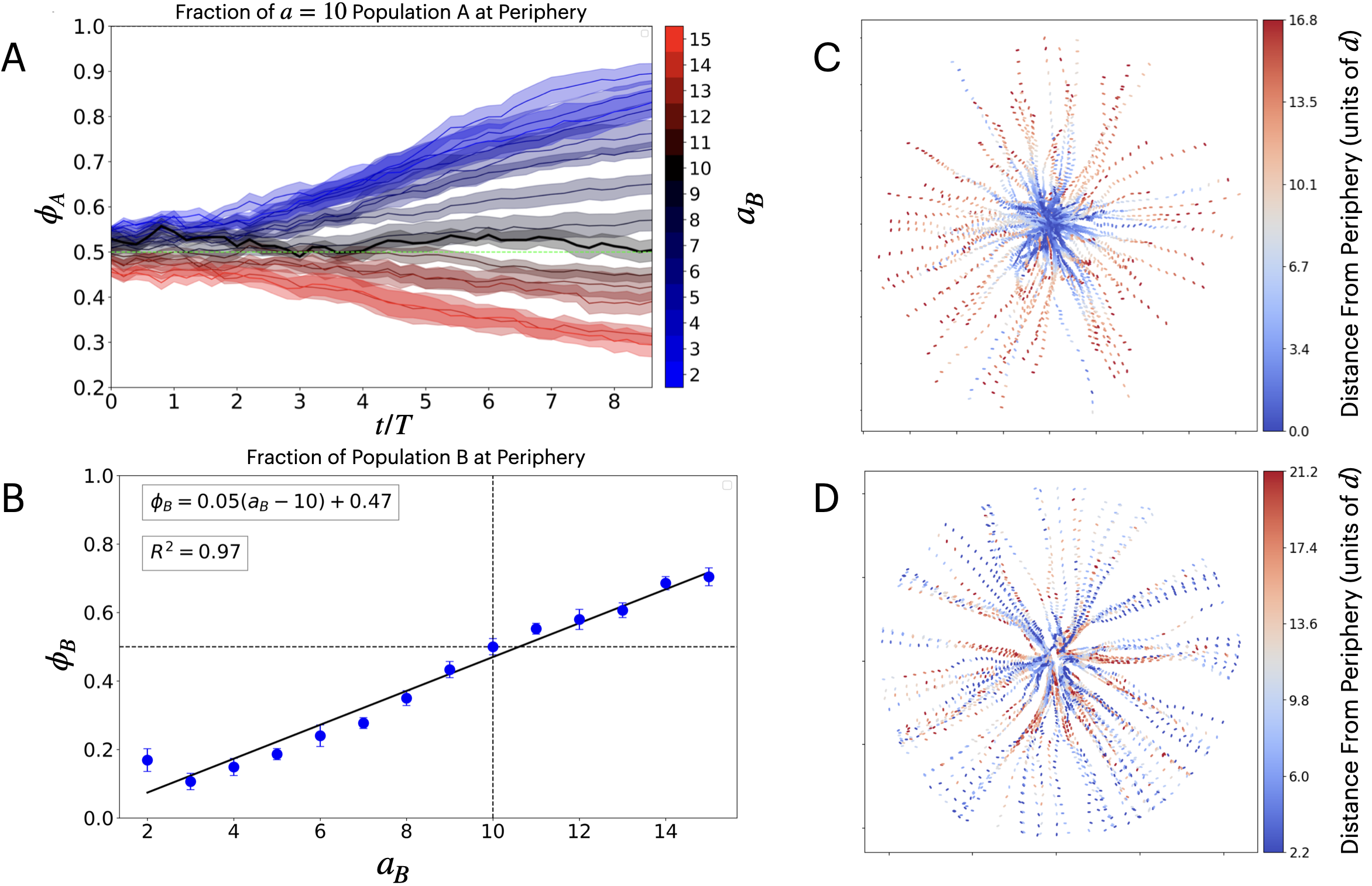}
    \caption{$\textbf{(A)}$ Time-series for the fraction $\phi_A$ of the colony periphery that is composed of Population A, which has fixed aspect ratio $a_A = 10$. The colony periphery comprises the cells whose centers lie within  a distance $5d$ from the computed alpha complex. Data is shown for various aspect ratios $a_B$ of Population B, each averaged over 10 independent simulations. \textbf{(B)} Same data as (A) but restricted to time $t=8.5T$, and here plotting the fraction $\phi_B=1-\phi_A$ of the periphery that is composed of Population B against its aspect ratio $a_B$. The black line is a linear best fit with fit parameters shown in the inset. Shaded regions in (A) and uncertainty bars in (B) represent standard error of the mean. \textbf{(C)} The lineage trajectories of randomly selected cells found in the bulk at the last time-step in a colony with $a_A = 5, a_B = 10$. \textbf{(D)} The lineage trajectories of randomly selected cells found in the periphery at the last time-step with $a_A = 5, a_B = 10$. In (C) and (D), trajectory points are colored by distance from the alpha complex at the time sampled in units of the cell width $d$, with red representing the deepest into the bulk.}
    \label{fig:fraction-periphery-timeseries}
\end{figure*}

To quantify the phenotypic spatial sorting, we investigate the statistics of the cells constituting the colony periphery. Figure~\ref{fig:fraction-periphery-timeseries}A displays the time series of the fraction of the periphery (by number of cells) made up by Population A cells, which we term the \textit{periphery fraction} of Population A. We define the periphery as those cells containing at least one point within a distance $w_p = 5 d$ of a piecewise-linear closed curve enclosing the colony called an alpha complex~\cite{Carlsson2024}, which is a generalization of the convex hull (see \textbf{Methods and Materials}). The periphery fraction $\phi_A$ of Population A increases over time when $a_A > a_B$ and decreases over time when $a_A < a_B$. This demonstrates that the sorting of large-aspect-ratio cells preferentially to the periphery is a gradually emergent phenomenon and not an artifact of initial conditions. We note that there is a small bias toward large-aspect-ratio cells at the periphery in the initial condition, a side effect of the initialization protocol (see \textbf{Methods and Materials}). While the equal-division-time rule $T_A=T_B$ gives greater total area to the longer cell type (Fig.~\ref{fig:growth}B), we emphasize that the number of cells of the two types remains approximately equal for the colony as whole, so $\phi_A \neq 0.5$ implies that the periphery has a different composition from the bulk.


We obtain a similar finding by examining the periphery fraction $\phi_B$ of Population B at the end of the simulation as a function of its aspect ratio $a_B$, with the aspect ratio of Population A held fixed at $a_A = 10$ (Fig.~\ref{fig:fraction-periphery-timeseries}B). The data reveals a strongly positive, linear correlation between $\phi_B$ and the aspect ratio $a_B$. This finding does not depend qualitatively on our choice of periphery width $w_p$ (Fig.~\ref{fig:changing_periphery_def}).

The importance of active mixing in generating turnover of cells at the periphery is illuminated by tracking the trajectories of selected cells and their ancestors backward in time. Some cells that are found deep in the colony bulk at final time (red) were on the periphery (blue) at earlier times (Fig.~\ref{fig:fraction-periphery-timeseries}C). Likewise, some cells that make up the final-time periphery begin in the colony bulk; they may even begin near the periphery, fall into the colony bulk, and return to the periphery (Fig.~\ref{fig:fraction-periphery-timeseries}D).  Together, these large variations in distance from the periphery over time indicate that substantial active rearrangements gradually alter the makeup of the periphery.

\begin{figure*}
    \centering
    \includegraphics[width=0.99\linewidth]{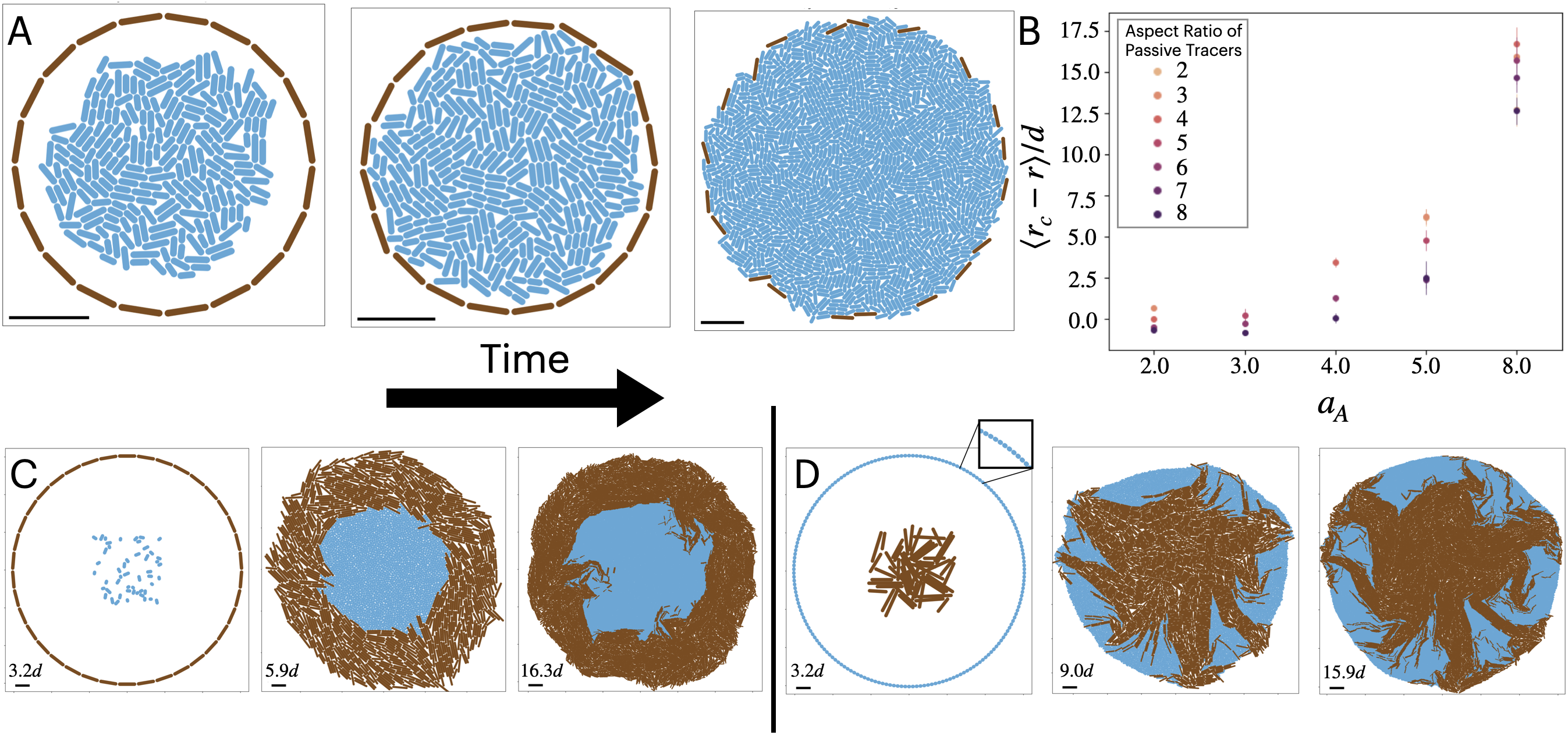}
    \caption{\textbf{(A)} Snapshots of the growth of a colony with  growing population A (blue, aspect ratio $a_A = 4$) initialized as an isotropic droplet inside a ring of non-growing,  passive tracers (brown, aspect ratio $a_P = 6$) with initially tangential orientations. The scale bars have length $20d$. \textbf{(B)} Mean radial displacement from colony periphery (radius $r_c$) of passive tracers at the final timestep, with the initial condition pictured in (A), for various aspect ratios $a_A$ and $a_P$. Colonies in (A) and (B) are grown to $10000$ cells in $5$ independent runs for each combination of parameters. Uncertainty bars represent standard error of the mean. \textbf{(C)} Snapshots of a colony simulated from an initial droplet of $N_A=50$ short cells (blue, $a_A = 2$) surrounded by a tangentially oriented ring of long cells (brown, $a_B = 10$), of radius $r = 3 \sqrt{N_A d^2 a_A / \pi }$ at times $t=0$, $t=5.2T$ and $t=8.28T$.
    \textbf{(D)} Snapshots of a colony simulated from an initial droplet of $N_B=50$ long cells (brown, $a_B = 10$) that is initially enclosed short cells (blue, $a_A = 2$) arranged in a ring of radius $r = 3 \sqrt{N_B d^2 a_B / \pi }$ at times $t=0$, $t=5.2T$ and $t=7.06T$.}
    \label{fig:annulus}
\end{figure*}

To further test the robustness of our finding that mechanical interactions preferentially sort longer-aspect-ratio cells to the periphery, we next examine whether cells of smaller aspect ratio are preferentially pulled away from the periphery. For this purpose, we simulated a single-population colony of Population A cells, initialized as an isotropic droplet, surrounded by a single-layer circle of passive tracers (non-growing rods), shown in Fig.~\ref{fig:annulus}C. We measure the inward radial displacement of these passive tracers relative to the periphery when the colony size reaches 10,000 cells. As shown in Fig.~\ref{fig:annulus}D, this radial displacement increases with increasing aspect ratio of Population A, presumably reflecting the enhanced active mixing seen in Fig.~\ref{fig:growth}D. Moreover, the radial displacement is consistently larger for smaller Population B aspect ratios. This trend indicates that larger cell aspect ratio accelerates the motion of the other cell type inward relative to the advancing periphery, leaving more of the periphery to be dominated by the longer cells. 

Replacing the circle of passive tracers with a circle of actively growing Population B in this scenario further confirms that cells of smaller aspect ratio tend not to reach the periphery (Fig. \ref{fig:annulus}A), whereas cells of longer aspect ratio often break through from the bulk to the periphery  (Fig.~\ref{fig:annulus}B). The latter seems to result in part from active mixing  that circulates smaller-aspect-ratio cells from the periphery into the colony bulk.

\subsection*{Radial Order}

In addition to the enhanced likelihood of larger aspect-ratio cells to be tangentially oriented at the periphery, Fig.~\ref{fig:annulus}B hints that large aspect-ratio cells have increased probabilities of being driven toward the periphery by internal pressures due to their higher local nematic order, as seen in Fig.~\ref{fig:defects}A-C. To test this hypothesis, we construct a global radial alignment parameter $\bar s_r$ that describes the radial or azimuthal alignment of cells within the colony. Following Ref.~\cite{Basaran2022}, we define $\bar s_r$ for a colony of $N$ cells as 
\begin{equation}\label{eq:radial_order}
    \bar s_r = \frac{1}{N}\sum_{i=1}^N \cos(2(\theta_i - \phi_i)),
\end{equation}
where $\theta_i$ is the angular orientation of the $i$th cell relative to a fixed axis, and $\phi_i$ is the azimuthal angle of the $i$th cell's position relative to the same axis and an origin at the colony's center of mass; the factor of 2 is introduced to account for the nematic head-tail symmetry of rod-like cells. With this definition, a colony of cells that are oriented along the radial direction (roughly the expansion direction) have $\bar s_r = 1$, while a colony of concentric circles will have $\bar s_r = -1$.

\begin{figure*}
    \centering
    \includegraphics[width=0.8\linewidth]{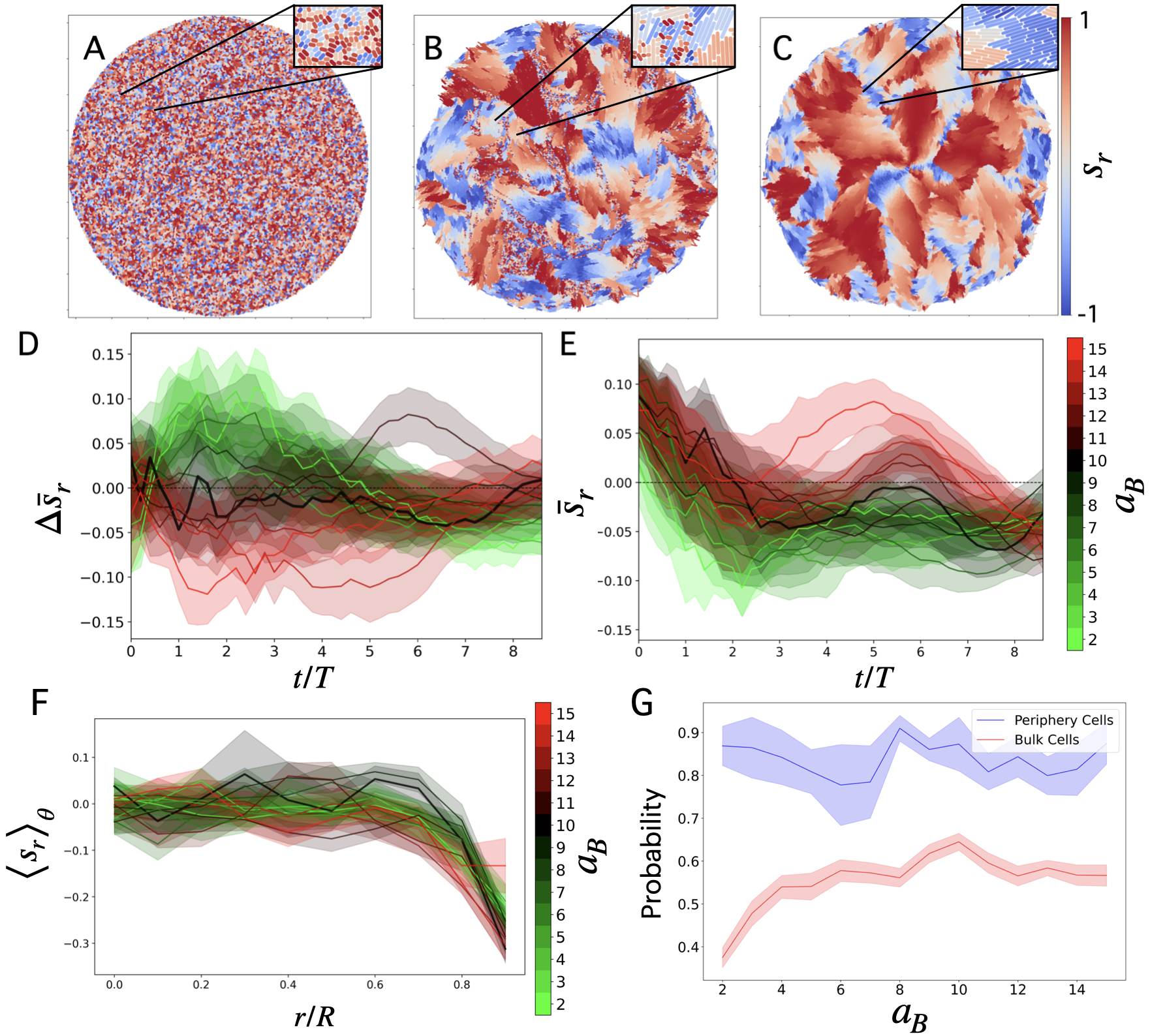}    
    \caption{\textbf{(A)-(C)} Snapshots of the final-time colonies from Fig.~\ref{fig:growth}A-C colored by the local radial order parameter $s_r$ (Eq.~\ref{eq:radial_order} with summation restricted to neighboring cells) for aspect ratios $(a_A, a_B)$ of \textbf{(A)} $(2,2)$, \textbf{(B)} $(2,10)$, \textbf{(C)} $(10, 10)$.  \textbf{(D)} Time-series of the difference $\Delta \bar s_r$ between the mean local radial order parameters $\bar s_{rA}$ and $\bar s_{rB}$ of Population A and Population B, respectively, with the aspect ratio of  Population A fixed at $a_A=10$. \textbf{(E)} Time-series of the mean radial order parameter $\bar s_r$ averaged over all cells.  \textbf{(F)} The azimuthally averaged radial order parameter $\left<s_r\right>_\theta$ at the final simulation time-step, plotted against distance $r$ from the colony center of mass in units of the radius $R$ of the circle circumscribing the colony. \textbf{(G)} Probability of periphery cells to have descended from a lineage with high radial order, using fixed $a_A = 10$ and varied $a_B$. The blue (red) line describes the probability of a periphery (bulk) cell, that was not at the periphery in the initial configuration, to come from a lineage that had radial order $s_r>0.8$ consecutively for two division times. Each colony is grown to 25000 cells and is initialized as an isotropic, well-mixed droplet of Population A and Population B in equal numbers. Uncertainty bands in (D, E, F, G) represent one standard error. }
    \label{fig:radial_panel}
\end{figure*}

We can further define a local order parameter $s_r(\mathbf{r})$ by constraining the averaging in Eq.~\ref{eq:radial_order} to a local neighborhood of cells that are in contact with a chosen cell. Figure~\ref{fig:radial_panel}A,B,C show final-time snapshots of the colonies from Fig.~\ref{fig:growth}A,B,C colored by $s_r(\mathbf{r})$. Comparison of $a_A = a_B = 2$ (Fig.~\ref{fig:radial_panel}A) and $a_A=a_B=10$ (Fig.~\ref{fig:radial_panel}C) shows that in colonies of uniform aspect ratio, larger aspect ratios produce larger regions of nematically ordered cells, consistently with previous findings~\cite{You2018}. Supplementary Movies S4, S5, and S6 show the growing colony colored by $s(r)$, corresponding respectively to Fig.~\ref{fig:radial_panel}A,B,C.

Cells at the periphery tend to be aligned tangentially, as seen in Fig.~\ref{fig:radial_panel}F; this trend has been characterized in prior studies as a form of active anchoring~\cite{Doostmohammadi2016,You2018}. In a bidisperse colony containing an equal mixture of  $a_A = 2, a_B=10$ cells (Fig.~\ref{fig:radial_panel}B), we find that the size distribution of nematically ordered domains shifts toward smaller sizes compared to the $a_A=a_B = 10$ case. 

To examine how radial order differs between the two cell types, we plot in Fig.~\ref{fig:radial_panel}D  a time-series for the difference $\Delta \bar s_r = \bar s_{rA} - \bar s_{rB}$ in the average radial order parameter between Population A (fixed $a_A=10$) and Population B ($a_B$ varied). In the early stages of the colony growth, this difference is positive for $a_B < a_A$ and negative for $a_B > a_A$, indicating that the more radially aligned cell type at early times is that with the higher aspect ratio. At later times, cells with larger aspect ratio instead have a higher tendency toward tangential alignment, due to their anchoring at the periphery and the buckling of radial domains in the colony bulk under active mixing. The colony-averaged radial order parameter $\bar s_{r}$ of both cell types together, plotted in Fig.~\ref{fig:radial_panel}E, shows an overall trend from preferentially radial to preferentially tangential alignment in the colony over time. At late times, the cells of larger aspect ratio anchor tangentially to the boundary, making the overall radial order parameter of the colony negative, as the bulk is isotropic (evidenced in Fig.~\ref{fig:radial_panel}F for small $r$). We note that the nonzero radial order parameter at $t=0$ is a consequence of our initialization procedure (see \textbf{Materials and Methods}).

As supporting evidence for the hypothesis that radially oriented nematic micro-domains are important to the eventual domination of the periphery by large-aspect-ratio cells, we examine whether cells in the final-time periphery are more likely than cells in the final-time bulk to have come from lineages with high radial alignment. We define a cell lineage as highly radial aligned if it maintains $s_r > 0.8$ for $2$ consecutive division times. For this measure, we exclude descendants of cells that were in the periphery of the colony's initial condition, and we randomly sample $1000$ cells from the bulk. Figure~\ref{fig:radial_panel}G shows this probability for a cell to have a history of high radial alignment, calculated for simulations with $a_A =10$ and $a_B$ varied. We find that cells that end in the periphery have a substantially higher probability of a history of high radial alignment, compared to cells that end in the bulk. The difference in radial history between the end-periphery and end-bulk cells is greatest when the aspect ratio of Population B is smallest.

\subsection*{Bidispersity Suppresses Active Mixing of Cell Types}
What consequences does bidispersity hold for genetic spatial structure? To address this question we now examine patterns in the heterozygosity, a measure of genetic diversity \cite{Korolev2010,Hallatschek2007,Farrell2017} that describes the probability of two sampled organisms to have distinct genotypes, as a function of spatial distance. In our model there is a one-to-one correspondence between genotype and phenotype, so heterozygosity quantifies how well-mixed are the two cell types. In particular, we examine the small-distance limit of heterozygosity, $H_0$, which is determined by sampling pairs of cells that are in contact; see \textbf{Materials and Methods}. Because our model has no mutation process, $H_0$ is zero inside a domain containing cells of a single type, and nonzero only on the boundaries between those domains or in well-mixed regions.

\begin{figure*}
    \centering
    \includegraphics[width=0.9\linewidth]{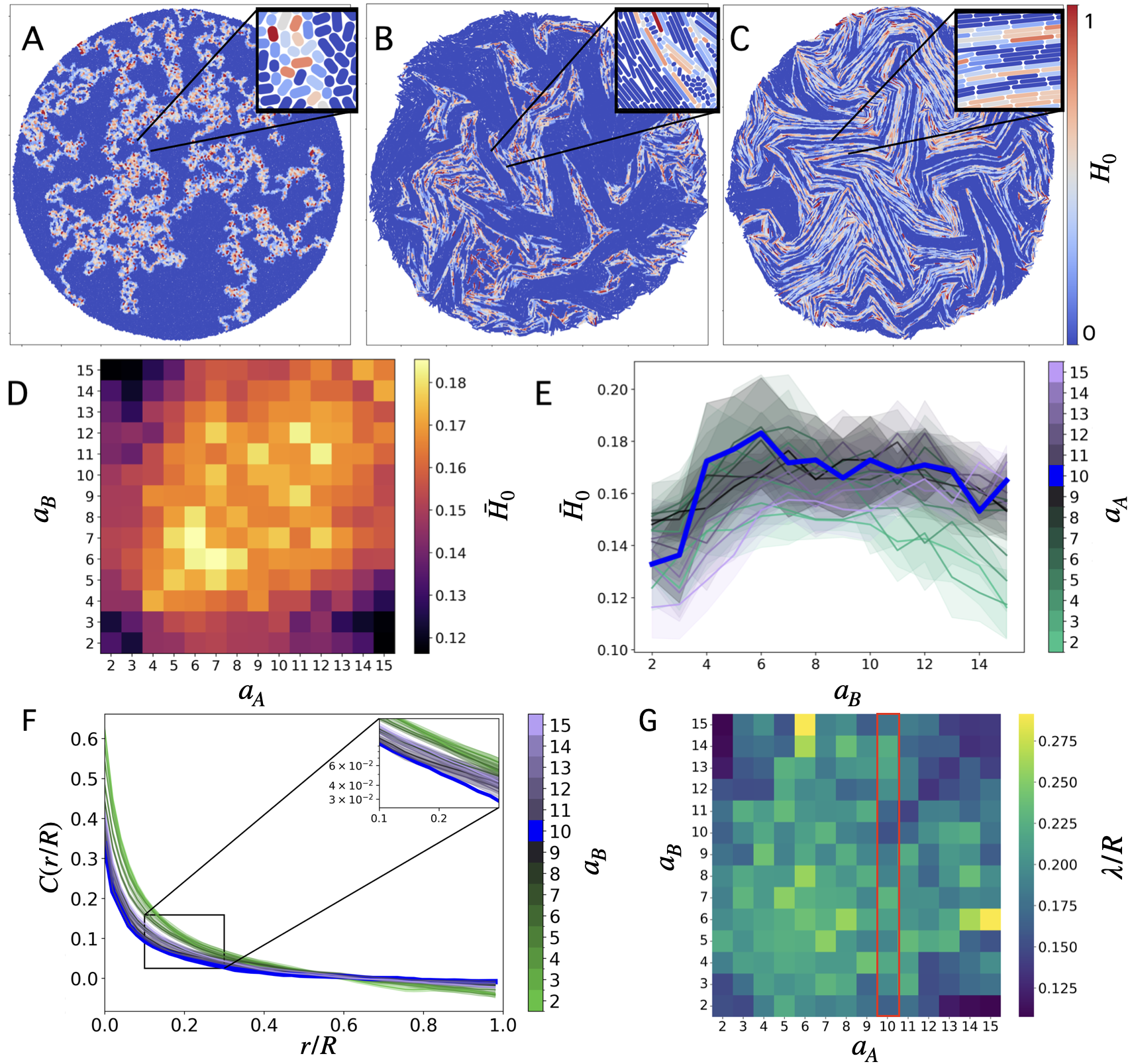}
    \caption{Snapshots of the final-time colony from Fig.~\ref{fig:growth}A-C colored by the local heterozygosity $H_0$ for aspect ratios $(a_A, a_B)$ of \textbf{(A)} $(2, 2)$,  \textbf{(B)} $(2, 10)$, \textbf{(C)} $(10, 10)$. \textbf{(D, E)} Final-time, colony-averaged local heterozygosity $\bar H_0$ for a range of aspect ratios $a_A$, $a_B$ of both sub-populations, plotted as a heatmap (D) and as line plots vs.\ $a_B$ (E). Data for equal aspect ratios $a_A=a_B$ are highlighted by the blue line in (E). \textbf{(F)} Final-time two-point cell-type correlation function $C(r/R)$, as a function of the ratio of the distance $r$ from the colony center of mass to the radius $R$ of the circle circumscribing the colony, for fixed $a_A=10$ and varied $a_B$. The inset is plotted with a logarithmic scale for $C$. Blue line highlights the equal aspect ratio ($a_A = a_B = 10$) case.  \textbf{(G)} Cell-type correlation length-scale $\lambda$ extracted from linear fits $-r/\lambda + C_0$ to $C(r)$ data. The column corresponding to the data in (F) is outlined in red. The fitting was restricted to $0.1 < r/R < 0.3$, the interval over which $C(r)$ is positive and approximately linear for all aspect ratios. Uncertainty bands in (E) and (F) represent one standard error.}
    \label{fig:heterozygosity_panel}
\end{figure*}

The local heterozygosity allows us to quantify the effect of cell aspect ratios on the degree of mixing, already hinted by the spatial structure shown in Fig.~\ref{fig:growth}. In Fig.~\ref{fig:heterozygosity_panel}A-C, we plot the colony snapshots of Fig.~\ref{fig:growth} with cells colored  by $H_0$. This measure of local genetic diversity is most suppressed when one or both cell types has small aspect ratio (Fig.~\ref{fig:heterozygosity_panel}A,B), due to the presence of large monoallelic clusters. Only when both cell types have large aspect ratio (Fig.~\ref{fig:heterozygosity_panel}C) does the colony avoid clusters by organizing most cells into thin bands of high nematic order, increasing the contact area between the two populations and thus making $H_0$ large in a greater fraction of the colony.

The influence of both cell aspect ratios on the heterozygosity at the end of the simulations is summarized in Fig.~\ref{fig:heterozygosity_panel}D, which plots $\bar H_0$, the local heterozygosity averaged over all cells in the colony, as a heatmap with $a_A$ and $a_B$ varied. $\bar H_0$ is maximal near the diagonal where the system is monodisperse, with $a_A=a_B$. Fig.~\ref{fig:heterozygosity_panel}E shows line plots of vertical sections of the data in Fig.~\ref{fig:heterozygosity_panel}D, with fixed $a_A$ and varied $a_B$ on each line. The blue line, representing the monodisperse case, appears to saturate for aspect ratios $a_A=a_B > 4$, consistent with calculations in Ref.~\cite{Schwarzendahl2022}, which also showed that nematic order occurs only in this regime. For the other, bidisperse cases, we find that mixing of the two cell types is significantly suppressed when one of them has small aspect ratio. When we hold one aspect ratio fixed at a small value $a_A \leq 3$, we find that mixing is minimized not only by making $a_B$ similarly small, as expected, but alternatively by making $a_B$ quite large, $a_B\gtrapprox 13$. Thus, extreme bidispersity seems to inhibit active mixing.

Further insight into the colony-scale structure of gene segregation is provided by the spatial two-point correlation function of cell type $C(r)$, which measures how the likelihood for two cells to have the same  cell type decays with distance. Figure~\ref{fig:heterozygosity_panel}F plots this decaying correlation function, with distance scaled by the colony radius $R$, using $a_A=10$ and various values of $a_B$. Over a modest interval $0.1R < r < 0.3R$, the data for all aspect ratio pairs are well-described by decaying exponentials, with correlation lengths $\lambda$ on the order of $0.2R$ (Fig.~\ref{fig:heterozygosity_panel}G). 
Smaller correlation lengths are obtained when one or both cell types have small aspect ratios, as well as when both types have large aspect ratios $a_A, a_B > 12$; the latter results from the cells' organization into thin, nematically ordered monoallelic bands, transverse to which cell type rapidly decorrelates. 
For the choice of $a_A=10$ in Fig.~\ref{fig:heterozygosity_panel}F, the monodisperse case $a_B=a_A$ has the lowest correlation, indicating the most mixing, over an interval $r \lessapprox 0.6 R$; however, we note that this finding is not generic to other choices of $a_A$. The slightly negative values of $C$ at larger $r$ are indicative of large clusters of the opposite cell type, and this behavior is generic to all $a_A$ values tested.

\subsection*{Equal Elongation Rate Growth}
\begin{figure*}
    \centering
    \includegraphics[width=0.9\linewidth]{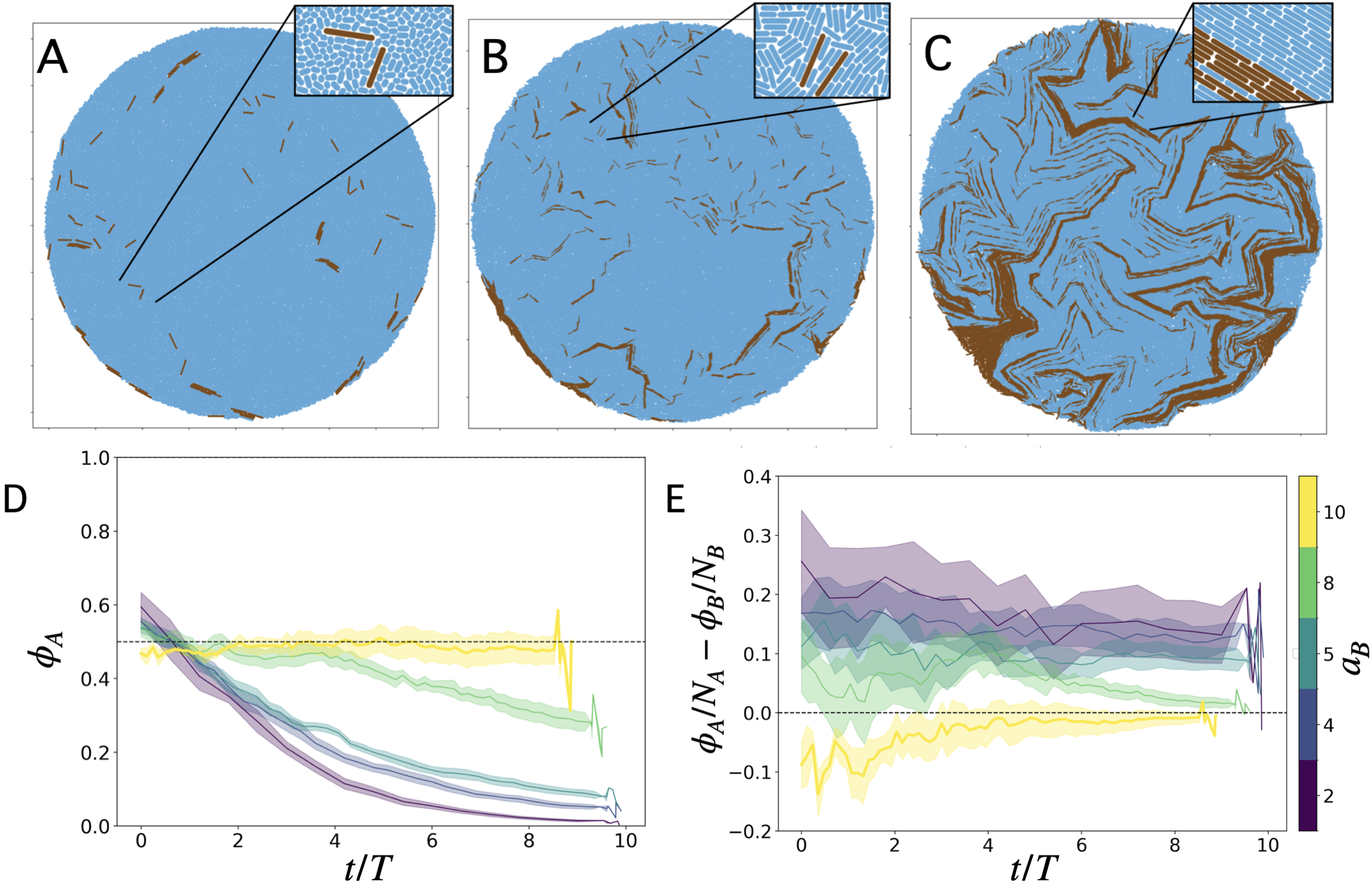}
    \caption{Snapshots of  final-time colonies grown with equal elongation rates $g_A=g_B$ for Population A (blue) and Population B (brown) of aspect ratios $(a_A, a_B)$ equal to  \textbf{(A)} $(2, 10)$, \textbf{(B)} $(5, 10)$, and \textbf{(C)}  $(8, 10)$. \textbf{(D)} Time-series for the fraction $\phi_A$ of the colony periphery composed of Population A ($a_A = 10$) for various aspect ratios $a_B$ of  Population B. The periphery is defined as the set of cells $\leq 5d$ away from the alpha complex.    \textbf{(E)} Time-series of the difference $\phi_A / N_A - \phi_B/N_B$, where $\phi_X$ is the periphery fraction and $N_X$ is the total number of cells for population $X$, with aspect ratio fixed at $a_A=10$ for A and varied for B. Each colony is grown to 25000 cells and is initialized as an isotropic, well-mixed droplet of Population A and Population B in equal numbers. Uncertainty bands in (D) and (E) represent one standard error. }
    \label{fig:equal_gr_panel}
\end{figure*}

To investigate how our findings depend on the choice to maintain equal numbers of population A and population B cells, we now examine the alternative growth protocol (2) in which the two cell types have equal elongation rate ($g_A = g_B = g= 1.5 \times 10^{-3} L_X/T_X$). The division times are now unequal, $T_A / T_B = L_A / L_B = a_A/a_B\equiv \alpha$. The number of A cells, $N_A \sim 2^{t/T_A}$, therefore grows with a different exponential rate compared to the number of B cells, $N_B \sim 2^{t/T_B}$, giving a number ratio $N_A / N_B \sim 2^{(t/T_A)(1 - T_A/T_B) } = 2^{(t/T_A)(1 - \alpha)}$. The ratio of areas occupied by the two populations therefore evolves as 
\begin{equation}
    \frac{A_A}{A_B} \sim \frac{d \cdot  L_A \cdot N_A}{d \cdot L_B \cdot  N_B} \sim \alpha \cdot 2^{(t/T_A)(1-\alpha)}.
\end{equation}

We perform five simulation runs and on a limited subset of aspect ratio combinations. The snapshots of the colonies after growth to  $25000$ cells are shown in Fig.~\ref{fig:equal_gr_panel}A-C.
As suggested by the growth snapshots, this means that the cell type with larger aspect ratio occupies an exponentially diminishing fraction of the colony area.
For this reason, the longer cells can no longer be said to dominate the periphery; their periphery fraction decays over time (Fig.~\ref{fig:equal_gr_panel}D). Nonetheless, we still find that cells of the larger aspect ratio are over-represented in the periphery compared to their prevalence in the bulk: The periphery fraction of population A is larger than the cell-number fraction of population A for the whole colony(\ref{fig:equal_gr_panel}E), meaning that population A is over-represented on the periphery, by an amount that increases as $a_A/a_B$ increases. 

\section*{Discussion}
The most striking feature of the equal division time growth protocol is the enhanced tendency of long cells to be located at the colony periphery compared to short cells.  The magnitude of this effect scales linearly with the ratio of aspect ratios of the two populations, as seen in Fig.~\ref{fig:fraction-periphery-timeseries}B. 
We hypothesize that the longer cells protrude through the periphery of the colony as a result of their likelihood of being contained within radially oriented nematic micro-domains. 

Note that this spatial segregation of cell types begins relatively early in the colony growth, as seen in Fig. \ref{fig:fraction-periphery-timeseries}A. This is consistent with radially ordered domains being more likely to form for the longer aspect ratio population at early times in the colony growth, as seen in Fig. \ref{fig:radial_panel}D. After around $t \approx 5 T$, these radial microdomains begin buckling and the longer aspect ratio cells begin reorienting themselves tangentially at the colony periphery.

However, the radially aligned regions do not appear to be solely responsible for the transport of larger cells to the periphery. In many cases they do not extend to the colony periphery and the radial order difference between $a_A =10 $ and $a_B$ in Fig.\ref{fig:radial_panel}D implies a higher radial order for the smaller-aspect-ratio cells at late times in the colony growth. The active mixing of the colony can slightly assist with spatial segregation, since the shorter cells can become mixed into the bulk from the periphery, as seen in Fig. \ref{fig:fraction-periphery-timeseries}B.
We conjecture that the emergent physics by which the longer rod-like cells are preferentially driven toward the periphery is analogous to granular convection\cite{koda1996smectic, adams1998entropically}, a common phenomenon in polydisperse granular systems whereby larger particles rise to the top of a shaken container \cite{rosato1987brazil,abreu2003influence}. 
Here, the ``top'' is the colony periphery, and the inward relative flow of short-aspect-ratio cells provides an analogous effect to gravity. 
Once at the periphery, the longer rods anchor tangentially to it and grow, thereby forming a fence-like barrier to cells in the interior. 

If our finding that the periphery is dominated by longer cells continues to hold true in scenarios where nutrients are not uniformly plentiful, we may then hypothesize that this physical phenotypic sorting phenomenon introduces an evolutionary selective pressure favoring elongated cell shapes: a mutation that increases a cell's aspect ratio will enable its progeny to become prevalent in the periphery, where nutrients are most abundant. The consequent increased growth rate would then create a positive feedback loop with increasing mutant fraction at the expansion front. A recent study \cite{Sreepadmanabh2024} suggests that larger-aspect-ratio bacteria also have an evolutionary advantage in constrained three-dimensional environments, which suggests that increasing cell aspect ratio might be a fitness-enhancing strategy under various growth conditions. 


\section*{Conclusions}
In this study we examined the spontaneous spatial sorting of cell types in a growing colony of rod-like bacteria with two different, heritable aspect ratios. We found that, increasingly over time, the colony periphery contains a disproportionately high number of the longer cells compared to their fraction of the total colony population. With the equal division time growth condition, we identified two mechanisms that favor the longer cells' abundance at the periphery. Firstly, the radial expansion pressure creates stable radial nematic microdomains for the longer cells, which we speculate enhances their ability to break through the periphery and thus join it. Secondly, longer cells promote active mixing dynamics which convects some cells from the periphery into the bulk. Tangentially oriented long cells, which are the least likely to be convected by these active mixing flows, come to occupy an increasing fraction of the periphery  over time. Furthermore, active anchoring seems to promote tangential alignment of the long cells at the periphery, which act in a fence-like manner to inhibit entrance into the periphery of the shorter cells from the bulk. In the alternative growth condition of equal elongation rates, we find that even though the shorter cells increase in number at a faster rate, the longer cells are nonetheless overrepresented at the periphery compared to the bulk.

We have also found that the growth of monoallelic domains, and the suppression of this growth by active mixing, depend on the aspect ratios of both populations, especially when one population is above the nematic threshold (aspect ratio $\approx 4$), and the other is below. This is due to the shorter-cell population forming large clusters, which are only slightly perturbed by thin nematic bands composed primarily of the longer-cell population.

Our findings suggest that it will be fruitful in future studies to examine how emergent active flows and shape segregation affect other types of interactions between microbial species. While explicit modeling of varying nutrient concentrations is outside the scope of this work, future simulation studies could couple bacterial growth to a nutrient field via a reaction-diffusion equation~\cite{Farrell2013,Farrell2017}, in order to test our hypothesis that active flows give longer cells an evolutionary advantage when nutrients are limited. Mutualistic or antagonistic interactions within bacterial communities, which respectively make mixing advantageous or disadvantageous, could also contribute to the evolutionary pressures affecting cell shape through the mechanisms explored here.  Additionally, spatial or temporal variation in nutrient availability could promote diversity in cell shapes through changes in the adaptive advantage offered by larger aspect ratio. Incorporating cell-scale mechanics and emergent active matter dynamics into these and other questions of population evolution will provide a more complete picture of the evolutionary dynamics of phenotypically diverse populations. 

\section*{Methods and Materials}
\subsection*{Simulation Initialization and Stopping Conditions}
Colonies are grown to a size of 25000 cells, which corresponds to approximately 9 generations of cell divisions ($t\approx 9T$), from an isotropic droplet of $50$ cells. Such a droplet is produced by initially placing cells of length $l_i= 0.01$ with uniformly distributed orientations and positions within a disk of radius $[50 d (L_A + L_B)]/4$. These cells are then grown to half their division size. This prevents overlaps in the initial cell configuration, at the potential cost of introducing a bias in orientations favoring the radial direction. Additionally, the shorter cells can more easily fit between longer cells, which makes them slightly more centrally localized on average.

\subsection*{Heterozygosity}
As a measure of the genetic structure and mixing in a two-phenotype system, we compute a variant of the heterozygosity $H$. This measure is useful in both experimental and theoretical contexts~\cite{Korolev2010,Hallatschek2007,Farrell2017} to quantify the probability that two randomly sampled constituents in a population will share the same phenotype. $H$ can depend on the sampling distance $r$ in a structured population~\cite{Korolev2010}. We examine the small sampling-distance limit $H_0$, which is proportional to the contact area between two sub-populations~~\cite{Schwarzendahl2022}.  Here, $H_0$ is computed through 
\begin{equation}
    H_0 = \frac{1}{N_T} \sum_i\left(\frac{1}{|N(r_i)|} \sum_{j \in N(r_i)} \delta (g_i - g_j)\right) \label{eq:h0}
\end{equation}
where $r_i$ is the position of the $i$th cell, $N_T$ is the total number of cells in the colony,  $N(r_i)$ is the set of indices enumerating the cells in contact with cell at position $r_i$. Here, $\delta(g_i - g_j) = 1$ if the cell types $g_i, g_j$ are equal, and is zero otherwise. 

For the equal elongation rate case, where the two cell types are not equal in number, $H_0$ as calculated by Eq.~\ref{eq:h0} will be biased toward zero since any region of fixed radius will contain a larger number of small aspect ratio cells compared to larger aspect ratio cells. One could alternatively calculate $H_0$ as
\begin{align}
    H_0 = \tfrac{1}{2} \left(H_{0}^A + H_{0}^B\right),
\end{align}
where $H_{0}^A, H_{0}^B$ are the average local heterozygosities restricted to each cell type, i.e. the average number of cells that cells of type A are in direct contact with cells of type B and \textit{vice versa}. 

\subsection*{Periphery Definition}
The boundary defining the colony periphery was computed using the Python library Alpha Shape Toolbox \cite{alphashape} with the endpoints of each rod (cell) used as the input for the alpha complex calculation. An edge between two points was included in the alpha complex if there was a disk of radius $1/\alpha$ that included both points on its boundary, but no other points inside. The value of $\alpha = 1.3$ was empirically chosen since it defines the colony boundary most precisely for our simulation runs, while still keeping it a single polygon. To determine which cells belonged to the periphery, a KD-tree was constructed from the points along the alpha complex, and the distance from each rod to the boundary was determined by calculating the nearest-neighbor distance for five equally spaced points along each rod. Cells located within a distance of $5d$ from the complex were classified as being on the periphery.
\FloatBarrier

\section*{Acknowledgments} 
Parts of this work were carried out at the Advanced Research Computing at Hopkins (ARCH) core facility  (rockfish.jhu.edu), which is supported by the National Science Foundation (NSF) grant number OAC1920103.

\printbibliography
\onecolumn
\section*{SI Appendix \label{sec:SI}}
\subsection*{Videos of the Growth}
\href{https://pages.jh.edu/dbeller3/resources/SI-Videos/Ratman-arXiv-2024/SuppVid1.mp4}{Supplementary movie S1}: Growth of a colony composed of cells with maximum aspect ratio $a_A=2$ (blue) and $a_B=2$ (brown), with equal division times $T_A=T_B$, from 50 to 25000 cells. 

\href{https://pages.jh.edu/dbeller3/resources/SI-Videos/Ratman-arXiv-2024/SuppVid2.mp4}{Supplementary movie S2}: Growth of a colony composed of cells with maximum aspect ratio $a_A=2$ (blue) and $a_B = 10$ (brown) with equal division times $T_A=T_B$, from 50 to 25000 cells. 

\href{https://pages.jh.edu/dbeller3/resources/SI-Videos/Ratman-arXiv-2024/SuppVid3.mp4}{Supplementary movie S3}: Growth of a colony composed of cells with maximum aspect ratio $a_A = 10$ (blue) and $a_B = 10$ (brown) with equal division times $T_A=T_B$, from 50 to 25000 cells.

\label{SM4} \href{https://pages.jh.edu/dbeller3/resources/SI-Videos/Ratman-arXiv-2024/SuppVid4.mp4}{Supplementary movie S4}: Growth of a colony composed of cells with maximum aspect ratio $a_A = 2$ and $a_B = 2$ with equal division times $T_A=T_B$, from 50 to 25000 cells, colored by radial order parameter. Red represents radially aligned cells, blue represents tangentially aligned cells. 

\label{SM5} \href{https://pages.jh.edu/dbeller3/resources/SI-Videos/Ratman-arXiv-2024/SuppVid5.mp4}{Supplementary movie S5}: Growth of a colony composed of cells with maximum aspect ratio $a_A = 2$ and $a_B = 10$ with equal division times $T_A=T_B$, from 50 to 25000 cells, colored by radial order parameter. Red represents radially aligned cells, blue represents tangentially aligned cells. 

\label{SM6} \href{https://pages.jh.edu/dbeller3/resources/SI-Videos/Ratman-arXiv-2024/SuppVid6.mp4}{Supplementary movie S6}: Growth of a colony composed of cells with maximum aspect ratio $a_A = 10$ and $a_B = 10$ with equal division times $T_A=T_B$, from 50 to 25000 cells, colored by radial order parameter. Red represents radially aligned cells, blue represents tangentially aligned cells.

\subsection*{Comparison of Various Definitions of the Periphery}

\begin{figure*}[h]
    \centering
    \includegraphics[width=0.97\linewidth]{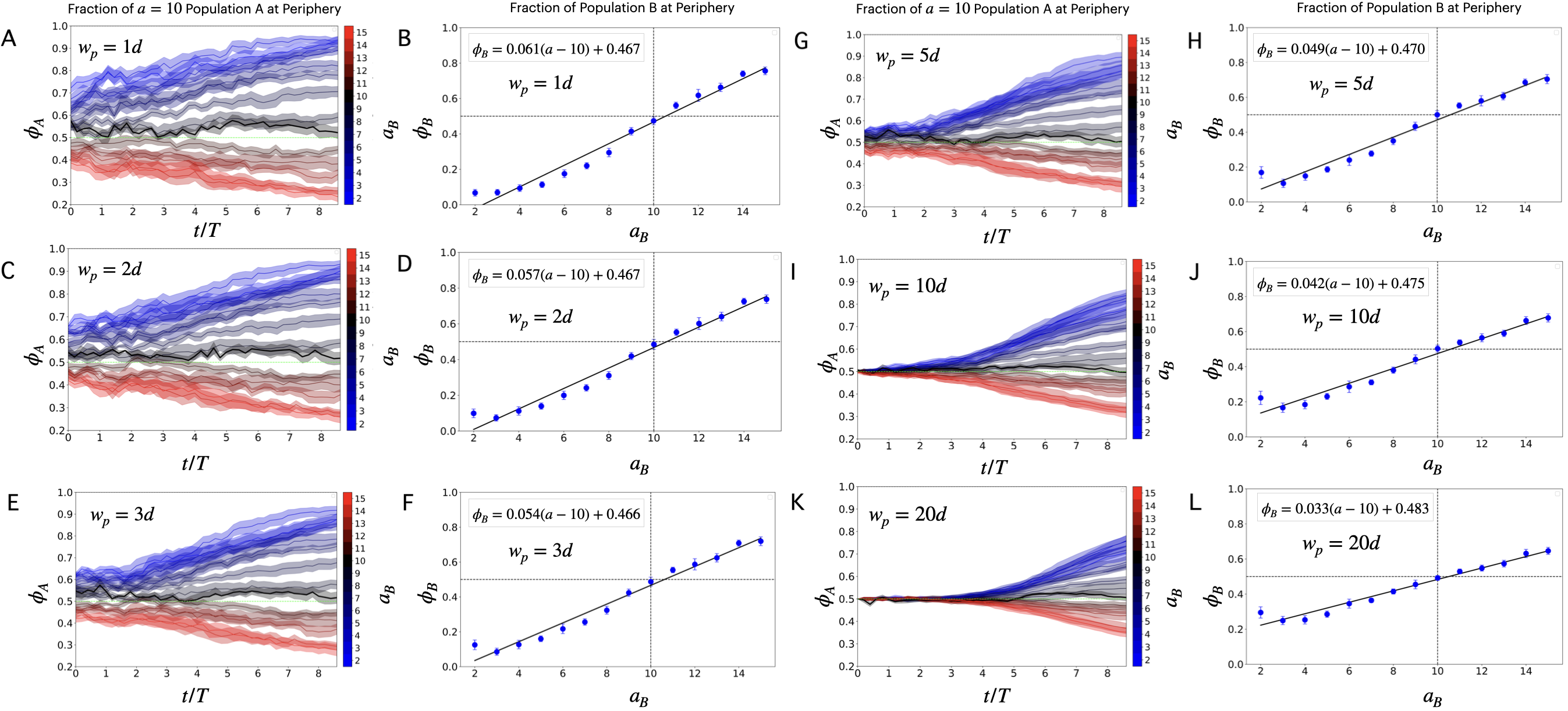}
    \caption{
    Effect of periphery thickness parameter on the measured phenotypic sorting. \textbf{(A), (C), (E), (G), (I), (K)} The fraction $\phi_A$ of $a_A=10$ cells at the periphery as a function of time for varying $a_B$ values. \textbf{(B), (D), (F), (H), (J), (L)} The fraction $\phi_B$ of Population B cells $a_B$ in the periphery plotted against that type's aspect ratio $a_B$.
    The periphery thickness parameter is 
    {(A, B)} $1d$, 
    {(C, D)} $2d$, 
    {(E, F)} $3d$,
    {(G, H)} $5d$,
    {(I, J)} $10d$,
    {(K, L)} $20d$, 
    in units of the cell width $d$. }    \label{fig:changing_periphery_def}
\end{figure*}

In Fig.~\ref{fig:fraction-periphery-timeseries} we define the periphery as those cells within a distance $w_p = 5d$ from the alpha complex. Here, we demonstrate that our findings are not qualitatively dependent on that choice of periphery thickness, and that the calculated preponderance of long cells decays as we increase the chosen periphery thickness parameter, as expected.  In Fig. \ref{fig:changing_periphery_def} we plot the periphery fraction of Population A for various choices of periphery thickness. For all choices of periphery thickness, there is a strong, positive linear correlation between the periphery fraction $\phi_B$ of Population B at the end of the simulation and its aspect ratio $a_B$ (Fig.~\ref{fig:changing_periphery_def}B,D,F,H,J,L), demonstrating that the phenotypic sorting effect is robust to this parameter choice. Furthermore, the slope of the $\phi_B$ vs.\ $a_B$ linear fit systematically decreases as the periphery thickness is increased, demonstrating that the dominance of long cells is a property of the periphery and not of the colony as a whole. We also observe that the phenotypic sorting effect appears to begin at later times when we use a larger periphery thickness (Fig.~\ref{fig:changing_periphery_def}A,C,E,G,I,K).


\subsection*{Nematic Order}
\begin{figure*}[h]
    \centering
    \includegraphics[width=0.9\linewidth]{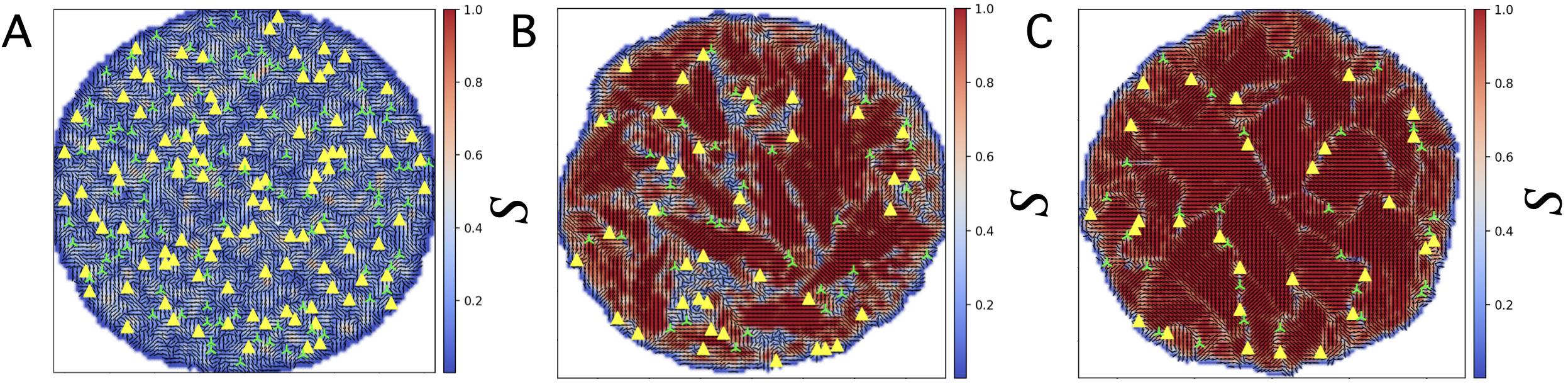}
    \caption{\textbf{(A)} \textbf{(B)}, \textbf{(C)} Topological defects ($+1/2$ yellow triangles and $-1/2$ green trefoils) and local nematic order $S$. The colonies have two sub-populations with aspect ratios $(a_A, a_B)$ equal to 
    {(A)} $(2,2)$, {(B)} $(2, 10)$, {(C)} $(10, 10)$. 
    In all subfigures, the colonies were grown to approximately 25000 cells.
    }
    \label{fig:defects}
\end{figure*}

Growing colonies of immotile, rod-like bacteria are known to exhibit short-range orientational order in nematic micro-domains  \cite{DellArciprete2018,You2018,Schwarzendahl2022}. To measure the degree of local nematic ordering, we calculate the nematic order parameter $S$, which is the largest eigenvalue of the nematic $\mathbf{Q}$ tensor \cite{mottram2014}, using uniform averaging over the cells whose centers lie within a disk of chosen cutoff radius $3(L_A + L_B)/8$ around each sampled point on a grid. We observe that the larger aspect ratio cells have higher local nematic order, as seen by comparing Fig. \ref{fig:defects}A and \ref{fig:defects}C. As mentioned in \textbf{Radial Order}, higher nematic order in the bulk aids longer cells in preferentially sorting to the periphery.

\end{document}